%% file: main.tex
\documentclass[runningheads]{llncs}

\usepackage[T1]{fontenc}
\usepackage{url} 
\usepackage{multirow}
\usepackage{multicol}
\usepackage{graphicx} 
\usepackage{amsmath}
\usepackage[utf8]{inputenc}
\usepackage{booktabs}
\usepackage{graphicx}
\usepackage{float}
\usepackage{adjustbox}
\usepackage{xcolor}
\usepackage{color}
\usepackage[varqu,scaled=0.88]{zi4}
\usepackage{listings}
\definecolor{pyctKeyword}{RGB}{0,73,114}
\definecolor{pyctString}{RGB}{128,64,16}
\definecolor{pyctComment}{RGB}{86,96,105}
\definecolor{pyctName}{RGB}{99,58,128}
\lstdefinestyle{pyctcode}{
  language=Python,
  basicstyle=\ttfamily\footnotesize,
  keywordstyle=\color{pyctKeyword},
  commentstyle=\color{pyctComment}\itshape,
  stringstyle=\color{pyctString},
  emph={math,range,len,sum,conc,register_indices},
  emphstyle=\color{pyctName},
  columns=fullflexible,
  keepspaces=true,
  showstringspaces=false,
  xleftmargin=0.4em,
  aboveskip=0.3em,
  belowskip=0.3em,
  breaklines=true
}
\usepackage{stmaryrd}
\usepackage{graphics}
\usepackage{epsfig}
\usepackage{colortbl}
\usepackage{makecell}
\usepackage{mathtools}
\usepackage[english]{babel}
\usepackage{amsfonts}
\usepackage{amsmath}
\usepackage{algorithm,algcompatible}
\algnewcommand\algorithmicreturn{\textbf{return}}
\algnewcommand\RETURN{\algorithmicreturn}
\algnewcommand\algorithmicprocedure{\textbf{procedure}}
\algnewcommand\PROCEDURE{\item[\algorithmicprocedure]}%
\algnewcommand\algorithmicendprocedure{\textbf{end procedure}}
\algnewcommand\ENDPROCEDURE{\item[\algorithmicendprocedure]}%
\algnewcommand{\algvar}[1]{{\text{\ttfamily\detokenize{#1}}}}
\algnewcommand{\algarg}[1]{{\text{\ttfamily\itshape\detokenize{#1}}}}
\algnewcommand{\algproc}[1]{{\text{\ttfamily\detokenize{#1}}}}
\algnewcommand{\algassign}{\leftarrow}

\newcommand{\OMIT}[1]{}
\newcommand{\rqtakeaway}[2]{%
  \par
  \vspace{0.35em}%
  \noindent\textbf{Takeaway (#1).} #2\par
  \vspace{0.35em}%
}

\begin{document}

\title{Influence-Guided Concolic Testing of Transformer Robustness}
\titlerunning{Influence-Guided Concolic Testing}

\author{Chih-Duo Hong \and Chih-Cheng Yang \and Yu Wang \and Fang Yu}
\authorrunning{C.-D. Hong et al.}

\institute{Department of Management Information Systems\\
National Chengchi University, Taipei, Taiwan\\
\email{\{chihduo,ccheng,ywang,yuf\}@nccu.edu.tw}\thanks{This research is supported by the National Science and Technology Council (NSTC), Taiwan, under grants 112-2222-E004-001-MY3 and 114-2634-F-004-002-MBK.}}

\maketitle

\begin{abstract}
Concolic testing for neural networks alternates concrete execution with
constraint solving to search for inputs that flip model decisions. We present
a concolic tester for Transformer classifiers that uses SHAP estimates to
rank pending path predicates by their impact on the current prediction. To
support self-attention with multiple heads in execution backed by SMT
solving, we implement attention semantics in pure Python that are compatible
with the solver and make the softmax boundary explicit by concretizing
exponentiation arguments. We evaluate our method on CIFAR-10
across three compact Transformer classifiers, ResNet18, and
VGG16 under a one-pixel budget and a 900\,s horizon. Across the 500
model--input pairs in this matched comparison, our method achieves 60\%
success, compared with 15\% for a differential evolution baseline that treats
the model as a black box. In the primary two-layer Transformer branch-ordering
study, SHAP-based predicate prioritization raises success from 56\% to 60\%
and reduces median attack time by 51\%. These results show that
influence-guided path exploration can make concolic testing a practical way
to find adversarial examples in recurrent and Transformer architectures.
\end{abstract}
\input{intro}

\input{related}
\input{method}
\input{testing}
\input{experiment_v2}

\section{Conclusion}
\label{sec:conclusion}

This paper presents a concolic tester for Transformer classifiers. The tester
uses SHAP scores to choose which pending branch predicate to solve next, and it
runs a Python version of attention that PyCT can track. The softmax step remains
approximate because exponentiation is concretized, so each candidate attack is
checked again on the original model. In the experiments, this design finds label
changes more efficiently than FIFO ordering and a matched differential evolution
baseline under the one pixel budget. These results support attribution as a
useful search signal, but not as a verification claim. Future work should reduce
solver and symbolic execution cost, extend the implementation to larger vision
transformers and richer tokenization pipelines, and study stronger perturbation
models and other priority rules.

\appendix
\section{Pure-Python Semantics of Multi-Head Attention with Concolic Execution}
\input{appendix}

\end{document}

%% file: intro.tex
\section{Introduction}

Robustness testing of neural classifiers is most useful when it produces
concrete, reproducible failures under a clearly stated perturbation budget.
Formal verification can provide stronger guarantees, but global robustness
claims are often too expensive or too restrictive for modern architectures
\cite{huang2020survey}. Attacks that treat the model
as a black box are easy to deploy, but they search without using the model's
internal path structure.
Concolic testing offers a middle ground: it executes the model concretely
while using SMT solving to explore alternate execution paths and synthesize
new inputs. Recent neural concolic testing work has used this idea for
adversarial example synthesis and fairness testing
\cite{yu2024constraintbased,huang2025concolic}. For robustness testing,
however, the value of concolic execution depends on whether the explored
paths are actually relevant to the model decision. This work builds on
PyCT~\cite{chen2021pyct}, a Python concolic testing framework that can track
symbolic expressions through ordinary Python operations.

This paper studies that problem for Transformer classifiers. Two
obstacles make the setting nontrivial. First, a neural network execution can
produce many symbolic branch alternatives, and solver time is easily spent on
predicates that are syntactically available but weakly connected to the
decision boundary. Prior testing tools for neural networks often rely on
coverage objectives such as neuron coverage or contribution
coverage \cite{sun2019deepconcolic}, but such structural criteria need not
align with behavior that changes the predicted label \cite{harel2020neuron}.
Recent Transformer-specific work includes attention-aware coverage
\cite{sekhon2022whitebox}, empirical ViT robustness and attacks
\cite{shao2022adversarial,jain2024towards}, and certified checking on
masking \cite{huang2023patchcensor}, softmax bounds \cite{wei2023convex}, or
linear relaxation encodings \cite{zhang2024galileo}.
These techniques either construct behavioral test suites, attack a fixed
model, or certify a specified perturbation region; they do not synthesize new
tests by executing a solver-compatible Transformer implementation and
negating observed path predicates. Second, standard Transformer
implementations rely on tensor libraries, softmax normalization, and attention
computations that do not expose the scalar Python control flow needed for
concolic execution in PyCT. Existing Transformer analyses that use constraints
therefore more often rely on certified Transformer encodings such as linear
relaxations or softmax bounds \cite{zhang2024galileo,wei2023convex} than on
synthesis from executable path constraints.

Our approach is to guide concolic testing by decision relevance while keeping its
validation boundary explicit. During execution, PyCT records bypassed branch
predicates. We associate each such predicate with the neurons affected by the
computation in which it arose, estimate those neurons' influence on the
current prediction using SHAP values, attribution scores derived from
SHapley Additive exPlanations \cite{lundberg2017unified}, and use the
resulting score to rank pending predicates. SHAP is therefore used
as a signal for controlling search, not as a substitute for solving: it changes only
which pending predicate is tried first. The path constraints themselves are
unchanged, and every reported label flip is checked by concrete execution of
the original model.

We also make the attention computation executable by the concolic engine.
Specifically, we implement self attention with multiple heads in pure Python for
PyCT, covering query/key/value projections, scaled dot product attention,
softmax over each row, value aggregation, and output projection using lists,
loops, scalar arithmetic, and ordinary conditionals. This is not a
mechanical port of a tensor program: the implementation must expose
scalar operations that generate branches while preserving enough correspondence
to the original model for concrete validation. Because exponentiation is not
directly supported by our SMT encoding, symbolic arguments for exponentiation
are concretized, yielding a symbolic approximation of softmax. Candidate
attacks synthesized through this executable semantics are validated by
running the original model on the resulting input.

We evaluate the approach on Transformer and CNN classifiers under a
one-pixel threat model. The experiments compare PyCT with a matched
differential evolution (DE) baseline~\cite{Su_2019} at a shared 900\,s
horizon, and separately compare SHAP ordering with an uninformed FIFO schedule
on the primary two-layer Transformer. The results show a broad paired advantage
over DE. They also show that influence-guided ordering lowers search cost and
modestly improves success, while the pure Python attention semantics makes
Transformer concolic execution feasible within the stated approximation.

Our main contributions in this work are as follows:

\begin{enumerate}
  \item \textbf{Concolic search guided by decision relevance.}
  We introduce a priority rule based on SHAP for concolic testing of neural networks
  that directs solver effort toward predicates associated with
  neurons relevant to the current decision rather than merely toward
  syntactically available branches or branches that increase coverage.

  \item \textbf{Executable attention semantics for PyCT.}
  We implement self attention with multiple heads as pure Python scalar computation,
  making attention paths visible to concolic execution while explicitly
  identifying the softmax approximation caused by concretized
  exponentiation.

  \item \textbf{Evidence under a matched robustness budget.}
  We evaluate the approach on compact Transformer classifiers under a shared
  budget that changes one input coordinate and show that it improves search cost
  relative to FIFO scheduling and finds more validated label flips than a
  matched DE baseline under the same one-pixel threat model.
\end{enumerate}

The rest of the paper is organized as follows. Section~\ref{sec:related-work}
reviews related work. Section~\ref{sec:methodology} presents the influence
guided testing method. Section~\ref{sec:transformer-semantics} describes the
attention semantics and toy concolic run. Section~\ref{sec:experiments}
reports the evaluation and limitations, and Section~\ref{sec:conclusion}
concludes. The appendix gives the expanded pure-Python attention semantics used
by the concolic executor.

\section{Related Work}
\label{sec:related-work}

Constraint-based analysis has long been used to reason about neural
networks \cite{huang2020survey}. Verification tools encode network
computations and safety properties as logical constraints; systems such as
Marabou focus on verification and analysis of deep networks
\cite{katz2019marabou}, while abstract domains and abstraction refinement
improve scalability for feedforward and recurrent models
\cite{singh2019abstract,lin2026robustness}.
These methods aim to prove or refute formal specifications. Our setting is
complementary: we use SMT solving inside a concolic loop to synthesize
concrete label-flipping inputs under an explicit perturbation domain.

Influence-guided testing connects our work to attribution methods. We use
DeepSHAP-style explanations \cite{lundberg2017unified} to estimate how much
intermediate neurons affect the current outputs. Unlike gradient-based attack
heuristics, which perturb inputs directly, our use of attribution is limited to
ordering pending symbolic branch alternatives during test generation.

Recent work has studied Transformer robustness primarily as an empirical
evaluation, attack, or defense problem. Shao et al.~\cite{shao2022adversarial}
report systematic studies of ViT robustness under adversarial perturbations,
and Jain and Dutta show that attention softmax can introduce gradient-masking
effects in ViTs \cite{jain2024towards}. These studies motivate robustness testing of
attention-based classifiers, but they evaluate models using gradient,
transfer, or benchmark attacks rather than by solving executable path
constraints.

Testing work for Transformers has adapted coverage criteria and model-based
checks to attention architectures. MNCOVER defines a white-box coverage
criterion for transformer-based NLP models that measures how attention-layer
behavior is exercised by a test suite \cite{sekhon2022whitebox}. Hong et al.
extract robust register automata as black-box surrogate models and use them to
assess robustness of recurrent and Transformer architectures
\cite{hong2025robustdra}. PatchCensor certifies patch robustness for ViTs by
exhaustively testing mutated attention masks and aggregating the resulting
predictions \cite{huang2023patchcensor}. These methods exploit Transformer
structure or learned surrogate models, but their objectives are test-suite
adequacy, model-based robustness assessment, or patch-defense certification. In
contrast, our tester records branch predicates during concolic execution and
uses those predicates to synthesize new candidate inputs under an explicit
perturbation domain.

Transformer robustness checking has moved beyond feedforward ReLU networks,
but it is still mostly based on relaxations, abstract interpretation, or
certification-specific encodings. Shi et al.~\cite{shi2020robustness} develop
robustness verification methods for self-attention and softmax, Bonaert
et al.~\cite{bonaert2021fast} improve certification precision for Transformer
models, and later work derives convex softmax bounds \cite{wei2023convex} or
tightens general linear relaxations for Transformers \cite{zhang2024galileo}.
These approaches seek proofs or certified accuracy for a specified
perturbation model. Our method is not a verifier: it uses SMT solving inside a
concolic loop to search for concrete label flips, ranks pending alternatives
with SHAP, and validates every reported candidate on the original model.

The gap is that these two lines of work leave different parts of the problem
uncovered. Transformer verification methods provide specialized encodings or
relaxations for attention and softmax \cite{shi2020robustness,zhang2024galileo},
whereas PyCT provides a Python-level concolic execution engine but not a
solver-compatible attention semantics \cite{chen2021pyct}. As a result,
attention operations, softmax normalization, and internal numerical branches
are not directly handled by existing neural concolic testers. Our contribution
is to connect these pieces in a narrower way: we give a concolic execution path
that makes compact attention classifiers analyzable within PyCT, while making
the softmax concretization and its validation boundary explicit.

%% file: method.tex
\section{Methodology}
\label{sec:methodology}

\subsection{Object-Oriented Concolic Testing}

Our tester builds on PyCT~\cite{chen2021pyct}, which exploits Python's dynamic object model rather than translating the program into a separate intermediate representation. PyCT replaces selected concrete values with \emph{concolic objects}. A concolic value is represented as a pair $\langle c, \phi \rangle$, where $c$ is the current concrete value and $\phi$ is a symbolic expression over the input variables. Operations on concolic values update both components: the original Python operation computes the concrete result, while the overloaded method constructs the corresponding symbolic expression whenever the operation is supported by the solver.

\begin{figure}[t]
  \centering
  \includegraphics[width=0.67\columnwidth]{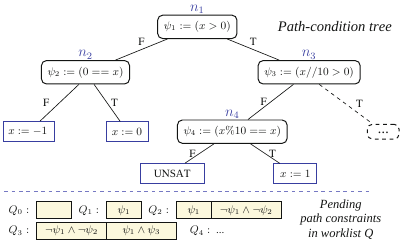}
  \caption{Example path-condition tree and path-constraint
  worklist in concolic execution. Internal nodes are branch
  predicates observed during execution, edges denote the concrete
  true/false outcomes, and the lower row shows snapshots of the
  worklist $Q$ containing unexplored path constraints. Each
  worklist entry preserves the executed prefix while negating one
  encountered branch predicate. Satisfiable entries yield new
  concrete inputs, while unsatisfiable entries are discarded.}
  \label{fig:path-tree}
\end{figure}

During one execution of the target function $f$, every branch
guard evaluated over concolic values contributes a literal to the
current path condition. At the same time, PyCT constructs
candidate alternatives for branches not taken during the current
execution. If the executed path contains taken branch literals
$\phi_1,\ldots,\phi_m$, then the alternative generated at depth
$i$ has the form
$
  \phi_1 \land \cdots \land \phi_{i-1} \land \neg \phi_i .
$
PyCT stores such formulas in a worklist $Q$, that is, a set of
pending path constraints that have not yet been submitted to the
SMT solver. Figure~\ref{fig:path-tree} illustrates the
path-condition tree and the corresponding updates to $Q$.
Solving a formula from $Q$ yields a new input intended to drive
the next execution through a previously unexplored branch.
The program is then re-executed concretely and symbolically,
and the process repeats until a counterexample is found or the
worklist or testing budget is exhausted.

In the original PyCT workflow, the worklist $Q$ can be processed by simple policies such as first-in-first-out. For neural-network testing, however, many predicates arise inside numerical computations, and these predicates are not equally relevant to the final model decision. The efficiency of the concolic search therefore depends heavily on which pending predicate is solved next. The next two subsections first define the influence-based score used for branch ranking and then show how this score is used in the concolic testing loop.

\subsection{Influence-Based Branch Scoring}

The concolic loop described above generates a queue of unexplored path
constraints. A default policy such as first-in-first-out treats all pending
constraints as equally useful. In neural-network testing, however, many
branches are introduced by internal numerical routines, and only a subset of
them is likely to affect the final prediction. We therefore replace the
ordinary queue with a priority queue whose ordering is derived from
SHAP-based neuron influence.

Let \(M\) be the neural model implemented by the executable Python function
\(f\), and let \(x_0\) be the seed input under test. During concolic execution,
each evaluated branch guard gives rise to an unexplored alternative: if the
current execution follows guard \(g\) under path prefix \(\pi\), the
alternative branch is represented by the formula \(\pi \wedge \neg g\), and
symmetrically for the other branch. We write \(b.\varphi\) for the formula
associated with such an unexplored branch \(b\).

To prioritize these branch formulas, we associate each unexplored
branch \(b\) with the neurons whose values are affected by the
computation in which the branch is evaluated. We denote this set by
\(\mathsf{Assoc}(b)\). This association is used only to rank pending
branches; it does not modify the path formula \(b.\varphi\) or the
constraint sent to the solver.

We now define the neuron-influence score used by this ranking.
Let \(a_l(x)\) denote the activation vector at layer \(l\), flattened
when the layer output is multidimensional, and let \(d_l\) be its
dimension. Let \(M_{>l}\) denote the suffix of the model from layer
\(l\) to the output layer, viewing \(a_l(x)\) as the input to this
suffix. For a neuron \(n\), let \(\ell(n)\) be the layer containing
\(n\), and let \(j(n) \in [d_{\ell(n)}]\) be the coordinate of \(n\)
in the flattened activation vector \(a_{\ell(n)}(\cdot)\).

Given a seed input \(x_0\) and a background dataset \(X\), define
for each layer \(l\)
\[
S_l \coloneqq M_{>l},~
z_l^0 \coloneqq a_l(x_0),~
B_l \coloneqq a_l(X) = \{a_l(x) : x \in X\}.
\]
Here \(S_l\) is the suffix model, \(z_l^0\) is the activation vector
of the seed input at layer \(l\), and \(B_l\) is a finite background
set in the activation space of \(S_l\). Thus, each
\(\tilde z \in B_l\) has the same dimension as \(z_l^0\).

We next define the SHAP attribution for a fixed layer \(l\). Let
\(z \in \mathbb{R}^{d_l}\) be an activation vector, let
\(B \subseteq \mathbb{R}^{d_l}\) be a finite background set, and let
\(o \in \mathcal{O}\) be an output coordinate, where \(\mathcal{O}\)
denotes the set of output neurons or classes. For
\(U \subseteq [d_l]\), define the hybrid activation
\(h_U(z,\tilde z)\) by
\[
h_U(z,\tilde z)_k \coloneqq
\begin{cases}
z_k, & k \in U,\\
\tilde z_k, & k \notin U.
\end{cases}
\]
Equivalently, \(h_U(z,\tilde z)\) keeps the coordinates in \(U\)
from \(z\) and fills the remaining coordinates
\(\bar U \coloneqq [d_l]\setminus U\) from the background activation
\(\tilde z\). We also write this hybrid vector as
\((z_U,\tilde z_{\bar U})\).
With \(S_l\) fixed, define
\[
v^{S_l}_{o,z,B}(U)
\coloneqq
\frac{1}{|B|}
\sum\nolimits_{\tilde z \in B}
\bigl(S_l(h_U(z,\tilde z))\bigr)_o .
\]
The notation \((\cdot)_o\) denotes projection onto the \(o\)-th
output coordinate. Thus,
\(\bigl(S_l(h_U(z,\tilde z))\bigr)_o\) is the logit, or output value,
of class/output neuron \(o\). The quantity \(v^{S_l}_{o,z,B}(U)\)
is therefore the empirical background expectation of the \(o\)-th
output when the coordinates in \(U\) are fixed to their values in
\(z\), while the remaining coordinates are marginalized using
background activations.

For an input coordinate \(j\) of \(S_l\), the corresponding SHAP
attribution to output \(o\) is
\[
\Phi_{j,o}(z;S_l,B)
\coloneqq
\!\!\sum\nolimits_{U \subseteq [d_l]\setminus\{j\}}
\!\!\!\frac{|U|!(d_l-|U|-1)!}{d_l!}\cdot \mathcal C(j,U),
\]
where $\mathcal C(j,U) \coloneqq v^{S_l}_{o,z,B}(U\cup\{j\})
-v^{S_l}_{o,z,B}(U)$ is the marginal contribution of
coordinate \(j\) when it is added to the coordinate subset \(U\).

The influence of neuron \(n\) at seed \(x_0\) is defined by
applying this layer-wise attribution at the layer containing \(n\):
\[
\mathsf{Shap}(n \mid x_0; M,X)
\coloneqq
\frac{1}{|\mathcal{O}|}
\sum\nolimits_{o \in \mathcal{O}}
\left|
\Phi_{j(n),o}
\bigl(
z_{\ell(n)}^0;
S_{\ell(n)}, B_{\ell(n)}
\bigr)
\right|.
\]
The absolute value treats both positive and negative effects on an
output as influence, and the average over \(\mathcal{O}\) avoids
tying the score to a single target class. In our implementation,
this quantity is estimated by the SHAP backend rather than by
enumerating all subsets \(U\). The definition above specifies the
attribution score that the priority heuristic approximates.

The influence of a branch is then the average influence of its associated neurons: $\mathsf{Infl}(b \mid x_0)$ is defined as
\[
\mathsf{Infl}(b \mid x_0) \coloneqq
\begin{cases}
\displaystyle
\sum\nolimits_{n \in \mathsf{Assoc}(b)}
\!\!\!\!\!\!\frac{\mathsf{Shap}(n \mid x_0; M,X)}{|\mathsf{Assoc}(b)|},
&
\mathsf{Assoc}(b)\neq \emptyset,\\[1.0em]
0,
&
\mathsf{Assoc}(b)=\emptyset.
\end{cases}
\]
The pure influence score can be used directly as the queue priority. In
larger models, however, highly influential branches may occur only after long
symbolic prefixes, causing expensive constraint construction and solver calls.
We therefore also use a depth-penalized priority score:
\[
\mathsf{Pri}(b \mid x_0)
=
(1-\alpha)\log_{10}\bigl(\mathsf{Infl}(b \mid x_0)+\epsilon\bigr)
-
\alpha \log_{10}(L_b+1),
\]
where \(L_b\) is the length of the path prefix of \(b\), \(\epsilon=10^{-12}\)
avoids taking the logarithm of zero, and \(\alpha\in[0,1]\) controls the
tradeoff between neuron influence and symbolic path length. Setting
\(\alpha\!=\!0\) yields a purely influence-driven ordering, while larger
\(\alpha\) values more aggressively penalize long path constraints.

\subsection{Influence-Guided Concolic Algorithm}
\label{sec:influence-guided-algorithm}

The scoring rule above determines only the order in which pending branch
alternatives are tried. Algorithm~\ref{alg:concolic_execution} shows the
resulting concolic loop: it keeps PyCT's concrete-symbolic execution and SMT
solving structure, but replaces the ordinary worklist with a priority queue
ordered by \(\mathsf{Pri}(b \mid x_0)\).

\begin{algorithm}[t]
\caption{Influence-guided concolic testing}
\label{alg:concolic_execution}
\begin{algorithmic}[1]
\State \textbf{Input:}
\State \quad Executable model \(f\), layered model \(M\), background dataset \(X\)
\State \quad Seed input \(x_0\), perturbation-domain predicate \(\Gamma(\mathbf{x};x_0)\)
\State \quad SHAP/path tradeoff coefficient \(\alpha \in [0,1]\)
\State \textbf{Output:}
\State \quad Counterexample \(x'\) such that \(x'\neq x_0\) and \(f(x')\neq f(x_0)\), or \(\bot\)
\State \textbf{Procedure:}
\State \quad \(\epsilon \gets 10^{-12}\)
\State \quad \(y_0 \gets f(x_0)\)
\State \quad \(x_{\mathit{cur}} \gets x_0\)
\State \quad \(I \gets \textsc{InnerShap}(M,X,x_0)\)
\State \quad \(Q \gets \textsc{PriorityQueue}()\)
\State \quad \textbf{repeat}
\State \quad\quad \((y,\mathcal{B}) \gets \textsc{ConcolicExec}(f,x_{\mathit{cur}})\)
\State \quad\quad \textbf{if} \(x_{\mathit{cur}}\neq x_0 \wedge y\neq y_0\) \textbf{then}
\State \quad\quad\quad \textbf{return} \(x_{\mathit{cur}}\)
\State \quad\quad \textbf{for each} \(b \in \mathcal{B}\) \textbf{do}
\State \quad\quad\quad \(s_b \gets 0\)
\State \quad\quad\quad \textbf{if} \(\mathsf{Assoc}(b)\neq\emptyset\) \textbf{then}
\State \quad\quad\quad\quad \(s_b \gets
        \operatorname{mean}\{I[n] : n\in \mathsf{Assoc}(b)\}\)
\State \quad\quad\quad \(L_b \gets \textsc{PathLen}(b)\)
\State \quad\quad\quad \(p_b \gets
        (1-\alpha)\log_{10}(s_b+\epsilon)
        - \alpha\log_{10}(L_b+1)\)
\State \quad\quad\quad \(Q.\textsc{push}(b,p_b)\)
\State \quad\quad \(x_{\mathit{next}} \gets \bot\)
\State \quad\quad \textbf{while} \(Q\neq\emptyset \wedge x_{\mathit{next}}=\bot\) \textbf{do}
\State \quad\quad\quad \(b \gets Q.\textsc{popMax}()\)
\State \quad\quad\quad \((r,\sigma) \gets
        \textsc{SolverCheck}\bigl(b.\varphi \wedge \Gamma(\mathbf{x};x_0)\bigr)\)
\State \quad\quad\quad \textbf{if} \(r=\textsc{sat}\) \textbf{then}
\State \quad\quad\quad\quad \(x_{\mathit{next}} \gets \textsc{ModelToInput}(\sigma)\)
\State \quad\quad \(x_{\mathit{cur}} \gets x_{\mathit{next}}\)
\State \quad \textbf{until} \(x_{\mathit{cur}}=\bot\)
\State \quad \textbf{return} \(\bot\)
\end{algorithmic}
\end{algorithm}

Relative to the baseline PyCT loop, the procedure differs mainly in the
scheduling of pending branch formulas. The solver still checks the same
symbolic branch alternative, but now under the perturbation domain
\(\Gamma(x;x_0)\), which encodes the allowed attack budget, fixed-coordinate
constraints, and input bounds. In the one-pixel experiments,
\(\Gamma(x;x_0)\) fixes all non-attacked input coordinates to their seed
values and constrains the selected normalized input coordinate to \([0,1]\).
Influence therefore changes which constraint is sent to the solver first, not
what the solver is asked to satisfy, so the ranking is a search heuristic
rather than a relaxation of the path condition. During the translated model's
forward computation, the implementation also records the neurons associated
with each symbolic computation: an output-neuron assignment is associated with
that singleton neuron, an intermediate value is associated with the output
neurons that depend on it, and a branch with no precise dependency set is
conservatively associated with the whole output layer. These associations are
used only for prioritization and do not alter concrete or symbolic execution.
For each seed input, the neuron-influence map is computed once before the
concolic search and cached for that seed by treating each layer activation as
the input to the suffix model and applying SHAP to estimate each neuron's
contribution to the final outputs. Algorithm~\ref{alg:shap-value-calculation}
summarizes this calculation.

\begin{algorithm}[t]
\caption{Inner-SHAP influence calculation}
\label{alg:shap-value-calculation}
\begin{algorithmic}[1]
\State \textbf{Input:}
\State \quad Model \(M\), background dataset \(X\), seed input \(x_0\)
\State \textbf{Output:}
\State \quad Influence map \(I:\mathcal{N}\rightarrow\mathbb{R}_{\geq 0}\)
\State \quad where \(\mathcal{N}\) is the set of layer-qualified neurons
\State \textbf{Procedure:}
\State \quad \(I \gets \emptyset\)
\State \quad \(\mathcal{O} \gets \textsc{OutputCoords}(M)\)
\State \quad \textbf{for each} non-output layer \(l\) of \(M\) \textbf{do}
\State \quad\quad \(S_l \gets M_{>l}\)
\State \quad\quad \(z_l^0 \gets a_l(x_0)\)
\State \quad\quad \(B_l \gets a_l(X)=\{a_l(x):x\in X\}\)
\State \quad\quad \textbf{for each} neuron \(n\) in layer \(l\) \textbf{do}
\State \quad\quad\quad \(j \gets j(n)\)
\State \quad\quad\quad \(I[n] \gets
    \displaystyle
    \frac{1}{|\mathcal{O}|}
    \sum\nolimits_{o\in\mathcal{O}}
    \left|
    \Phi_{j,o}
    \bigl(
    z_l^0; S_l, B_l
    \bigr)
    \right|\)
\State \quad \textbf{return} \(I\)
\end{algorithmic}
\end{algorithm}

During search, each concolic execution returns the set \(B\) of unexplored
branch alternatives encountered on the current path. Each record contains the
complete path formula \(b.\varphi\) and associated-neuron set
\(\mathsf{Assoc}(b)\); the tester assigns priority
\(\mathsf{Pri}(b \mid x_0)\), inserts \(b.\varphi\) into the priority queue,
and repeatedly asks the solver for the highest-priority satisfiable formula. A
satisfying assignment is converted into the next concrete input and executed
again under concolic tracking. With finite time and memory budgets, the
procedure terminates when it finds a counterexample, exhausts the priority
queue, or reaches the global budget. Without such budgets, the search is not
complete because the number of feasible paths can grow exponentially. Every
reported counterexample is nevertheless validated by concrete re-execution of
the original model: Algorithm~\ref{alg:concolic_execution} returns \(x'\) only
after observing \(f(x')\neq f(x_0)\) in an actual execution. Thus,
influence-guided scheduling may miss counterexamples, but it does not
introduce spurious label flips.

%% file: testing.tex
\section{Testing the Transformer Architecture}
\label{sec:transformer-semantics}
\label{sec:running}

\begin{figure}[t]
    \centering
    \includegraphics[width=0.8\linewidth]{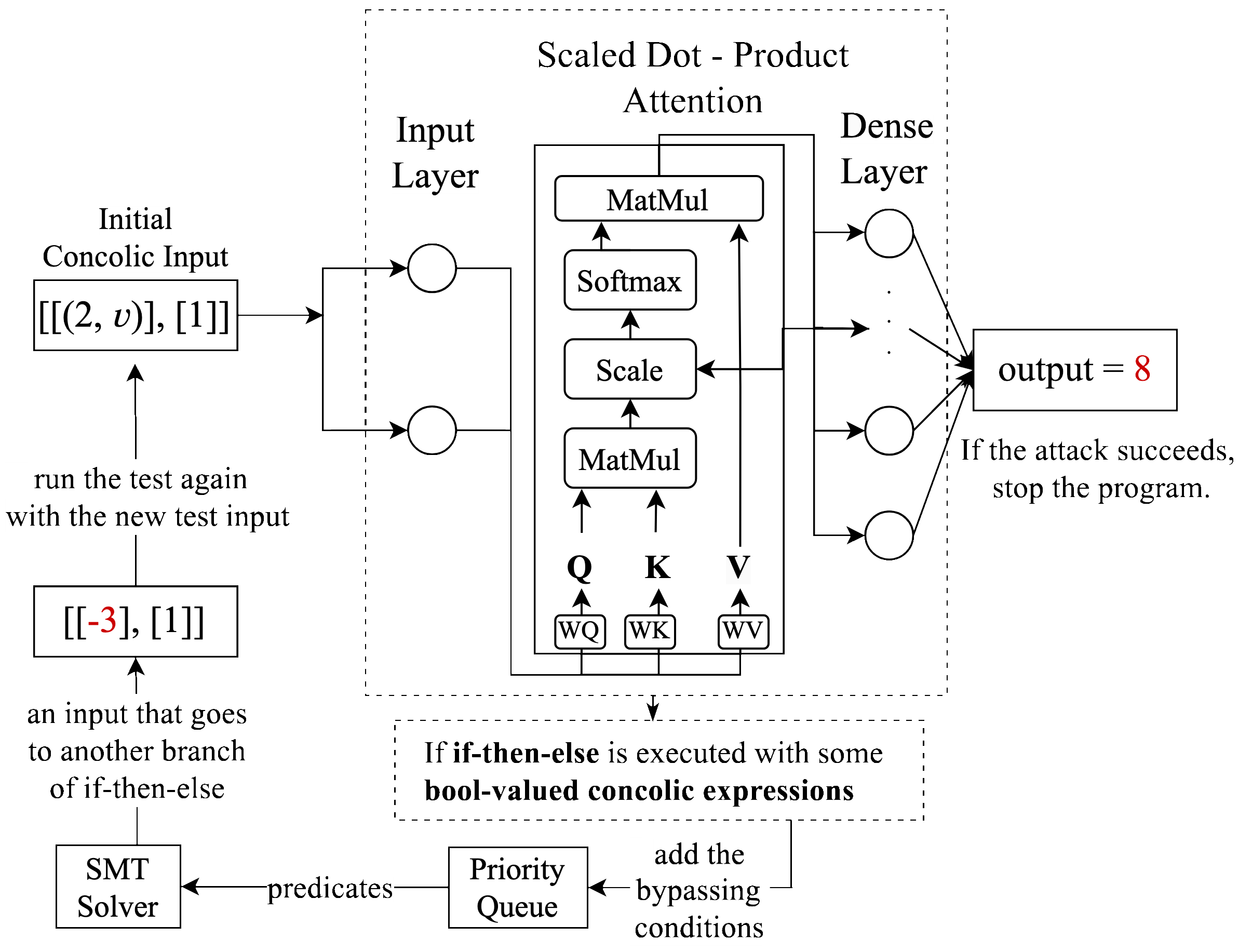}
    \caption{A concolic run on a toy single-layer Transformer. PyCT executes
the pure-Python attention block, records bypassed branch predicates from the
softmax computation, prioritizes them using SHAP-based influence, solves one
predicate, and re-executes the model on the resulting concrete input.}
    \label{fig:concolic_testing}
\end{figure}

PyCT operates on plain Python objects and control flow, but modern
Transformer implementations usually rely on tensor libraries that do not
expose concolic semantics. We therefore translate the Transformer forward
pass into Python lists, loops, scalar arithmetic, and conditional branches.
 
\subsection{Solver-Compatible Attention Semantics}

For an input sequence $X\in\mathbb{R}^{L\times d_{\mathrm{model}}}$, each
attention head $i$ computes
\[
Q_i=XW^Q_i\!+\!B^Q_i,\quad
K_i=XW^K_i\!+\!B^K_i,\quad
V_i=XW^V_i\!+\!B^V_i,
\]
\[
S_i[t,u]=\!\sum\nolimits_j \!\frac{Q_i[t,j]K_i[u,j]}{\sqrt{d_k}}\!,~
P_i=\mathrm{softmax}(S_i),~
A_i=P_iV_i.
\]
The head outputs are concatenated and projected by
\[
Y[t,\ell]=\sum\nolimits_i\sum\nolimits_u A_i[t,u]W^O[i,u,\ell]+B^O[\ell].
\]
The implementation realizes these operations using Python lists, loops,
scalar arithmetic, and ordinary conditionals. Branch predicates arise from
Python boolean tests over concolic scalars; in the multi-head attention
block, the dominant source is the row-wise \texttt{max} loop used by the
numerically stable softmax. For solver compatibility, arguments to
\texttt{math.exp} are concretized:
\[
P_i[t,u]=
\frac{\exp(\mathrm{conc}(S_i[t,u]-m_{i,t}))}
{\sum\nolimits_v \exp(\mathrm{conc}(S_i[t,v]-m_{i,t}))}.
\]
Thus, symbolic execution tracks the scalar comparisons that select the row
maximum, while the exponential terms are evaluated with concrete arguments.
Candidate attacks are still checked by concrete re-execution of the original
model.

\input{example}

\OMIT{
The main source of branch predicates in this block is the row-wise
\texttt{max} loop inside softmax. When a concolic guard $g$ is evaluated,
PyCT records the taken predicate in the current path condition and queues the
negation of the taken predicate as an unexplored alternative. Thus, if
\texttt{z > m} evaluates true, the queued alternative is $z \le m$; if it
evaluates false, the queued alternative is $z > m$. For the toy weights used
in Figure~\ref{fig:concolic_testing}, the seed value $v=2$ yields concrete
scores satisfying
\[
S[0,1] < S[0,0]
\quad\text{and}\quad
S[1,1] < S[1,0].
\]
The current execution therefore takes the false branch in both comparisons
and queues the bypassed alternatives
\[
\varphi_1:\; S[0,1] > S[0,0]
\;\equiv\;
6v+6 > 3v^2+6v+3,
\]
\[
\varphi_2:\; S[1,1] > S[1,0]
\;\equiv\;
12 > 6v+6.
\]
If \(\varphi_2\) is selected, the solver may return $v=-3$, yielding the
concrete input $X'=[[-3],[1]]$. Re-executing the original model on $X'$
takes an alternate softmax-max branch and changes the prediction from label
$9$ to label $8$, so the attack is reported as successful. If re-execution
does not change the label, PyCT continues with the next queued predicate.
}

\subsection{Computing SHAP-Based Neuron Scores}
\label{sec:shap-score-computation}

The priority rule in Section~\ref{sec:influence-guided-algorithm} requires one scalar influence score
for each tracked neuron position. We compute these scores once for each
seed input \(x_0\) before concolic search begins, cache them by
layer-qualified tensor position, and use the cache only to prioritize
pending branch predicates.

\paragraph{Sequential neural models} For sequential models, we compute
layer-wise attributions by treating the remaining network suffix as the
model to be explained. At layer \(l\), the seed and background set are
propagated to that layer, giving \(z_l^0 \coloneqq a_l(x_0)\),
\(B_l \coloneqq a_l(X)\), and suffix model \(S_l \coloneqq M_{>l}\). We apply the
\emph{gradient explainer}~\cite{lundberg2017unified} to \(S_l\), using
\(B_l\) as the background distribution and \(z_l^0\) as the explained
activation. The returned attribution tensor is averaged over output
coordinates, and over the batch dimension when present, to obtain one
score per neuron position. The procedure then advances one layer and
repeats on the next suffix. This slicing is well defined for sequential
networks because the computation has a single feed-forward path.

\paragraph{Functional, DAG, or residual models}
For functional, DAG, and residual models, layer-by-layer slicing can
break skip connections, shared tensors, or merge operations. We
therefore preserve the full computational graph. The input layer still
uses standard SHAP values, while intermediate layers use a
graph-preserving influence proxy. A feature model returns all tracked
intermediate activations together with the final logits. Let \(T\) be
the scalar target logit for this seed, namely the unique output logit in
the binary case or the current top-class logit in the multi-class case.
For tracked activation coordinate \(i\) in layer \(l\), we compute
\[
I_l(i)
=
\frac{\partial T}{\partial a_l(i)}
\left(
a_l(x_0)_i
-
\frac{1}{|X|}\sum\nolimits_{x\in X} a_l(x)_i
\right).
\]
This value is large when the activation differs from its background
baseline and the target logit is locally sensitive to that activation.
Unlike exact SHAP, this proxy does not enumerate coalitions of hidden
neurons and does not claim the Shapley axioms for DAG intermediate
layers. Its role is only to provide a stable, low-cost ordering signal
while preserving the original model graph.

Both computation paths expose the same interface to the concolic tester:
each tracked neuron position receives one scalar influence value.
Downstream branch prioritization is therefore independent of whether the
value came from layer-wise SHAP or from the graph-preserving proxy. In
both cases, the score is only a search heuristic; all reported
counterexamples are still validated by concrete re-execution of the
original model.

%% file: example.tex
\subsection{Example of a Concolic Execution Run}

Figure~\ref{fig:concolic_testing} illustrates one concrete concolic run on a
toy single-layer Transformer. The seed input is $X_0=[[2],[1]]$, which is
classified as label $9$ in this illustrative toy classifier. PyCT promotes
the first scalar to a concolic value,
$X=[[\llbracket 2,v\rrbracket],[1]]$,
where $2$ is the concrete value used in the current execution and $v$ is the
symbolic variable submitted to the solver.
For the weights used in this example, we have
\[
W_Q=[[[1,1]]],\quad W_K=[[[2,1]]],\quad W_V=[[[1,2]]],
\]
\[
B_Q=[[1,1]],\quad B_K=[[2,1]],\quad B_V=[[1,2]],
\]
and the symbolic components of the projected matrices are
\[
\begin{aligned}
Q=\left[\begin{smallmatrix}
v+1 & v+1\\
2 & 2
\end{smallmatrix}\right],\quad
K=\left[\begin{smallmatrix}
2v+2 & v+1\\
4 & 2
\end{smallmatrix}\right],\quad
V=\left[\begin{smallmatrix}
v+1 & 2v+2\\
2 & 4
\end{smallmatrix}\right].
\end{aligned}
\]
Thus the scaled dot-product score matrix is
\[
S=\tfrac{1}{\sqrt{2}}
\left[\begin{smallmatrix}
3v^2+6v+3~ & 6v+6\\
6v+6 & 12
\end{smallmatrix}\right].
\]
The relevant code path in this example is row-wise softmax. The code below
omits tensor details and keeps only the operations that matter for
concolic execution.

\vspace{.5em}
\begin{lstlisting}[style=pyctcode]
def softmax(scores):
    row_maxima = [row_max(scores, t) for t in range(len(scores))]
    exp_rows = [[math.exp(conc(value - row_maxima[t]))
                 for value in scores[t]] for t in range(len(scores))]
    return [[v / sum(row) for v in row] for row in exp_rows]

def row_max(scores, t):
    idx = [(t, k) for k in range(len(scores[t]))]
    register_indices(idx)
    current = scores[t][0]
    for u in range(1, len(scores[t])):
        if scores[t][u] > current: current = scores[t][u]
    return current
\end{lstlisting}
\vspace{.5em}

The call to \texttt{register\_indices}
records that the subsequent guards in \texttt{row\_max} belong to row
\(t\) of the attention-score matrix. The actual symbolic branch predicates
come from the Python guard \texttt{scores[t][u] > current}: when either side is
concolic, PyCT records the guard taken in the current execution and queues
the negated guard, together with the same path prefix, as an unexplored
branch alternative. The row index recorded by \texttt{register\_indices}
lets the branch handler associate that queued predicate with the
corresponding attention-row positions and assign a SHAP-based priority. The
\texttt{conc} call in \texttt{softmax} denotes the concrete projection used
before \texttt{math.exp}, so in this example the symbolic branch alternatives
come from the \texttt{row\_max} comparisons rather than from the exponential
computation.

At the seed value \(v=2\), the concrete scores satisfy
\[
S[0,1] < S[0,0]
\quad\text{and}\quad
S[1,1] < S[1,0].
\]
Hence the current execution keeps the first entry as the maximum in both rows. The bypassed true branches are queued as candidate path predicates:
\begin{align*}
\varphi_1:\; S[0,1] > S[0,0]
& \;\equiv\;
6v+6 > 3v^2+6v+3,\\
\varphi_2:\; S[1,1] > S[1,0]
& \;\equiv\;
12 > 6v+6.
\end{align*}
The priority queue orders such predicates by their SHAP-based influence scores. If \(\varphi_2\) is selected, the solver may return \(v=-3\), which yields the concrete input $X'=[[-3],[1]].$
PyCT then re-executes the model on \(X'\). In the example shown in Figure~\ref{fig:concolic_testing}, this input takes an alternate branch in the softmax maximum computation and changes the model prediction from class \(9\) to class \(8\). The attack is therefore reported as successful. If the solved input does not change the predicted label, PyCT continues by selecting another predicate from the queue.

%% file: experiment_v2.tex
\section{Experiments}
\label{sec:experiments}

Our evaluation is organized around three research questions.

\begin{itemize}
\setlength{\itemsep}{0.2em}
\setlength{\parsep}{0pt}
\item \textbf{RQ1.} Does PyCT find more one-pixel attacks that change
the model label than a matched differential evolution baseline that
treats the model as a black box
under the same attack budget?
\item \textbf{RQ2.} Does SHAP-guided branch ordering reduce the cost
of concolic search compared with an uninformed queue policy?
\item \textbf{RQ3.} How sensitive is PyCT to the SHAP/path-length
tradeoff, and can the gains be explained by coordinate selection
alone rather than symbolic solving?
\end{itemize}

We answer RQ1 with a comparison across models at a matched horizon, and answer RQ2 and RQ3
using the primary two-layer Transformer.
All experiments were conducted on a desktop computer with an R7 7700 CPU and 64GB RAM.

\subsection{Experimental Setup}
\label{sec:exp-setup}

\paragraph{Datasets and models} Our evaluation uses five CIFAR-10 classifiers: three compact Transformer models and two CNN baselines. The Transformer with one layer reshapes the $32 \times 32 \times 3$ input into $1024$ tokens of dimension $3$, applies one multi-head attention (MHA) block, and then uses a dense classifier head. The primary model is the two-layer attention classifier. It adds a second attention stage after a dense bottleneck and reshape, and produces a latent token representation of shape $16 \times 8$. The variant with eight blocks makes this design deeper by stacking several MHA blocks over two stages. We also include ResNet18 and VGG16 to test whether the same search heuristics work on CNNs.

\paragraph{Attack workflow} We evaluate the first 100 CIFAR-10 test samples in index order, without filtering out inputs that are already misclassified. SHAP-guided runs use a fixed background set of 30 test images, selected by stratified sampling with three images per class and sampling seed 2233. Under the one-pixel budget, PyCT promotes only the selected normalized input coordinate to a concolic variable, initialized at its clean value in the \([0,1]\) model input domain. The main controls are the attack mode, timeout and solver allocation, the SHAP/path coefficient \(\alpha\), and the symbolic-state threshold; all PyCT runs use CVC5. Because we do not have separate translation-equivalence measurements, a candidate is counted only after concrete re-execution on the original model. Reported PyCT times include concolic execution, constraint construction and solving, and validation, but exclude SHAP computation because that runtime was not logged.

Under prioritization based on SHAP, each pending branch receives a score that balances local influence and symbolic search depth:
$$
\mathrm{score}(\alpha) \;\coloneqq\; (1-\alpha)\log_{10}(|s|+\varepsilon)-\alpha\log_{10}(L+1),
$$
where \(s\) is the SHAP value of the corresponding branch position, \(L\) is the current path length, and \(\varepsilon=10^{-12}\) prevents taking the logarithm of zero. Thus, branches with higher scores are explored first. Smaller \(\alpha\) places more weight on SHAP influence, whereas larger \(\alpha\) more aggressively penalizes long symbolic paths.

\paragraph{Baseline} We compare against a differential evolution (DE) baseline \cite{Su_2019}. This baseline treats the model as a black box. It directly optimizes a perturbation vector with low dimension and observes only model outputs; it does not build symbolic formulas or use the internal model structure. Both tools are evaluated under the same one-pixel threat model, on the same first 100 test cases, and under the same time budget. An attack is counted as successful when the model's top-1 prediction changes relative to its original prediction on the clean input. This comparison separates symbolic reasoning that uses internal model structure from a generic search that treats the model as a black box.


\subsection{Overall Comparison under a One-Pixel Attack Budget}
\label{sec:exp-overall}

\begin{figure}[!t]
  \centering
  \begin{minipage}[t]{0.32\textwidth}
    \centering
    \includegraphics[width=\linewidth]{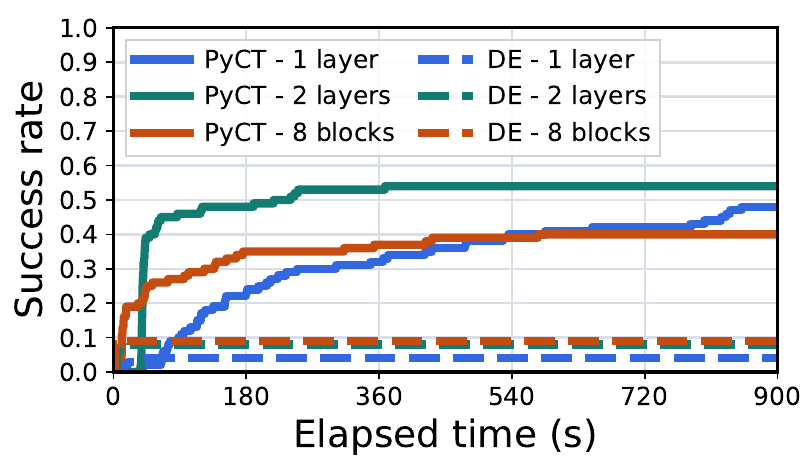}
    \centerline{\small (a) Transformer family}
  \end{minipage}
  \hfill
  \begin{minipage}[t]{0.32\textwidth}
    \centering
    \includegraphics[width=\linewidth]{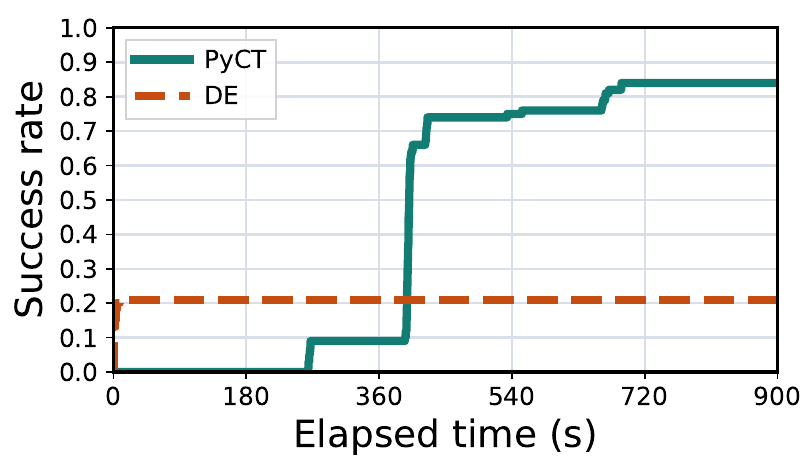}
    \centerline{\small (b) ResNet18}
  \end{minipage}
  \hfill
  \begin{minipage}[t]{0.32\textwidth}
    \centering
    \includegraphics[width=\linewidth]{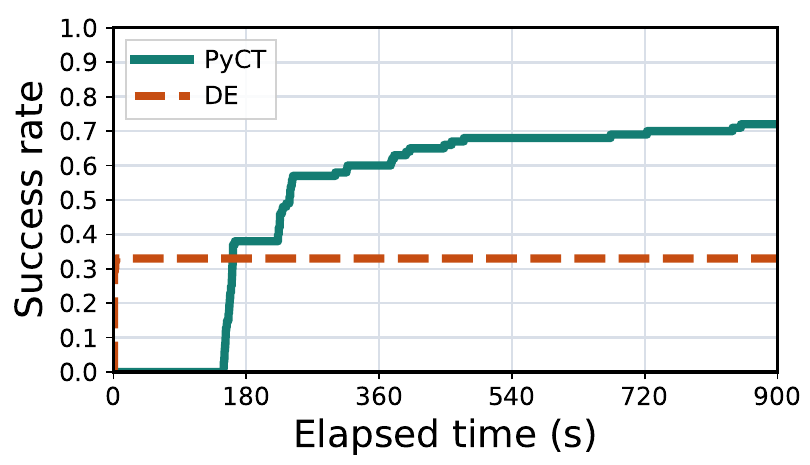}
    \centerline{\small (c) VGG16}
  \end{minipage}

  \caption{Cumulative attack rates under a one-pixel budget, comparing PyCT with the DE baseline over time.}
  \label{fig:tool-comparison-all}
\end{figure}

To answer RQ1, Figure~\ref{fig:tool-comparison-all} summarizes the comparison between PyCT and the DE baseline. The first plot combines the three Transformer models. The other two plots show ResNet18 and VGG16. Together, the panels show how quickly each tool finds successful attacks under the same one-pixel budget and time limit.

The cumulative curves in Figure~\ref{fig:tool-comparison-all} show the main trend. Under the shared 900\,s horizon, PyCT finds successful attacks earlier than the DE baseline for all model families. Table~\ref{tab:overall-matched-summary} gives the numerical gap.

\begin{table}[!t]
\centering
\caption{Comparison across models under a fixed 900\,s budget. \emph{Params.} gives the number of trainable model parameters; \emph{DE succ.} and \emph{PyCT succ.} report top-1 label change success rates; \(\Delta\)succ. is the PyCT--DE difference in percentage points; the mean columns report average runtime in seconds. All PyCT runs in this table use the same one-pixel threat model, fixed \(\alpha=0.6\), and 900\,s timeout. The symbolic state threshold is chosen for each model family.}
\label{tab:overall-matched-summary}
\footnotesize
\setlength{\tabcolsep}{3pt}
\renewcommand{\arraystretch}{0.96}
\begin{tabular*}{0.98\textwidth}{@{\extracolsep{\fill}}lrrrrrr@{}}
\toprule
Model & Params. & \makecell{DE\\succ.} & \makecell{PyCT\\succ.} & \makecell{$\Delta$\\succ.} & \makecell{DE\\mean} & \makecell{PyCT\\mean} \\
\midrule
Transformer (1 layer) & 395,609 & 4 & 46 & +42 & 867.2 & 469.8 \\
Transformer (2 layers) & 396,769 & 8 & 54 & +46 & 829.6 & 272.0 \\
Transformer (8 blocks) & 108,262 & 9 & 40 & +31 & 819.7 & 327.2 \\
ResNet18 & 11,183,562 & 21 & 85 & +64 & 711.8 & 500.3 \\
VGG16 & 14,995,146 & 33 & 75 & +42 & 603.4 & 439.1 \\
\bottomrule
\end{tabular*}
\end{table}

The DE curves also show a clear plateau near the end of the 900\,s horizon. In that region, additional elapsed time does not substantially increase the baseline success rate. This behavior is consistent with a population search that observes only model outputs and has already used its configured query budget, or has stagnated on the remaining inputs. After that point, more time does not necessarily lead to new model queries or restarts. We therefore interpret the plateau as a search stagnation or evaluation budget effect, not as evidence that the remaining inputs are robust. PyCT is different because symbolic branch exploration can still expose new candidate paths, so a longer time limit can lead to more validated label changes.

\begin{table}[t]
\centering
\caption{Paired comparison on the same first 100 CIFAR-10 test cases under the shared 900\,s horizon. \emph{Both succeed} counts inputs for which both methods change the label; \emph{PyCT only} and \emph{DE only} count inputs attacked successfully by only one method; \emph{Neither} counts inputs where both methods fail. 
}
\label{tab:overall-paired}
\footnotesize
\setlength{\tabcolsep}{3pt}
\renewcommand{\arraystretch}{0.96}
\begin{tabular*}{0.82\textwidth}{@{\extracolsep{\fill}}lrrrr@{}}
\toprule
Model & \makecell{Both\\succeed} & \makecell{PyCT\\only} & \makecell{DE\\only} & Neither \\
\midrule
Transformer (1 layer) & 1 & 45 & 3 & 51 \\
Transformer (2 layers) & 7 & 47 & 1 & 45 \\
Transformer (8 blocks) & 5 & 35 & 4 & 56 \\
ResNet18 & 20 & 65 & 1 & 14 \\
VGG16 & 31 & 44 & 2 & 23 \\
\bottomrule
\end{tabular*}
\end{table}

Table~\ref{tab:overall-paired} gives a paired view of the same first 100 inputs. The main pattern is the large gap between \emph{PyCT only} and \emph{DE only}. Across all five models, PyCT succeeds on 236 inputs where the DE baseline fails within the shared 900\,s horizon. The reverse happens on only 11 inputs. Thus, the margin in Table~\ref{tab:overall-matched-summary} reflects a broad advantage across inputs rather than a small number of outlier cases.

Aggregating the discordant outcomes across all 500 model--input pairs,
PyCT succeeds where DE fails in 236 cases, whereas DE succeeds where
PyCT fails in 11 cases. An exact two-sided McNemar test,
equivalently a binomial test on the 247 discordant pairs under the null
hypothesis that either method is equally likely to be the unique
success, gives $p = 3.9 \times 10^{-56}$. Wilson 95\% confidence intervals
for the aggregate success rates are $60.0\%$ [$55.6\%$, $64.2\%$] for
PyCT and $15.0\%$ [$12.1\%$, $18.4\%$] for DE.

The overlap also differs by model family. On the Transformer models, the two methods overlap on relatively few successful cases (1, 7, and 5 inputs). This suggests that PyCT finds many successful attacks that the baseline does not reach under the same budget. On ResNet18 and VGG16, the overlap is larger (20 and 31 inputs), but the paired comparison still strongly favors PyCT.

\rqtakeaway{RQ1}{Under the matched one-pixel and 900\,s horizon, PyCT
shows a broad paired advantage over the DE baseline: it
succeeds on 300/500 model and input cases versus 75/500 for DE, with 236
successes by PyCT alone and only 11 successes by DE alone.}

\subsection{Primary Model Study: Two-Layer Transformer}
\label{sec:exp-transformer}

We now turn to the primary two-layer Transformer. This model is a useful setting for studying the influence-guided design. It is large enough to show nontrivial symbolic growth, but still compact enough for systematic sweeps over the search parameters.


\paragraph{Branch ordering and internal baselines}
To answer RQ2 and isolate the effect of branch ordering, we compare \texttt{queue} and \texttt{shap} under identical conditions: the same two-layer Transformer, test cases, one-pixel budget, SHAP-selected symbolic pixel, timeout, solver wrapper, and symbolic path threshold. The only difference is how pending constraints are ordered for exploration. Under these matched settings, SHAP-guided ordering improves success modestly but reduces search cost substantially, especially median attack time and explored path length. In other words, SHAP guidance does not just find more attacks; it finds shorter and cheaper symbolic paths.

This comparison also helps explain the failures. Queue order spends more solver calls on long symbolic prefixes and leaves more cases unresolved by timeout. SHAP-guided order converts a small number of those timeouts into successful attacks and reaches the remaining unresolved cases with lower cost. The improvement is therefore not only a final count effect; it is also visible in the search cost profile.

\rqtakeaway{RQ2}{SHAP-guided ordering primarily improves search
efficiency: on the two-layer Transformer, it raises success from 56/100
to 60/100 while reducing median attack time from 467.7\,s to 230.1\,s
and median path length from 746 to 285.}

\begin{table}[!t]
\centering
\caption{\(\alpha\) sweep on the two-layer Transformer under solver timeout \(=60\)\,s, attack timeout \(=1800\)\,s, and path-length threshold \(=2000\). Success and Timeout count cases out of 100; Mean time and Med.~time report mean and median attack time in seconds; Med.~succ. is the median time among successful attacks; Path len. is the median path length.}
\label{tab:two-layer-alpha}
\footnotesize
\setlength{\tabcolsep}{3pt}
\renewcommand{\arraystretch}{0.96}
\begin{tabular*}{0.88\textwidth}{@{\extracolsep{\fill}}rrrrrrr@{}}
\toprule
\(\alpha\) & Succ. & Timeout & Mean time & Med.~time & Med.~succ. & Path len. \\
\midrule
0.0 & 40 & 60 & 1201.4 & 1800.1 & 73.8 & 1029 \\
0.2 & 53 & 47 & 920.6 & 463.1 & 39.2 & 804 \\
0.4 & 58 & 42 & 830.6 & 232.1 & 37.2 & 456 \\
0.6 & 60 & 40 & 799.4 & 230.1 & 37.9 & 285 \\
0.8 & 60 & 40 & 813.7 & 267.7 & 37.4 & 256 \\
1.0 & 60 & 40 & 829.1 & 238.5 & 37.8 & 258 \\
\bottomrule
\end{tabular*}
\end{table}

\paragraph{Effectiveness of the SHAP score}
To answer the sensitivity component of RQ3, Table~\ref{tab:two-layer-alpha}
reports the latest settings of the
SHAP vs.~path-length tradeoff coefficient $\alpha$. Recall that the
branch priority is computed by $\mathrm{score}(\alpha)$,
where small $\alpha$ emphasizes estimated decision influence,
while large $\alpha$ focuses on symbolic tractability by penalizing long paths.

The sweep shows that pure influence is too expensive, and that the best
success/time tradeoff occurs at an intermediate setting. Pure influence
can select highly impactful branches that occur deep in the attention
or softmax computation, where long symbolic prefixes are expensive
to solve. The pure path-length setting reaches the same success count as
$\alpha=0.6$, but with higher mean and median attack time; it favors
cheap predicates without directly checking whether they are relevant to
the label decision. The intermediate setting $\alpha=0.6$ gives the
best balance: it ties for the highest success count while giving the
lowest mean and median attack time among the top-success settings.

This pattern also explains why SHAP-guided scheduling mainly improves
search cost rather than changing the feasible search space. SHAP does
not change the perturbation domain or relax the path constraints; it
only changes the order in which pending constraints are explored.
By moving decision-relevant predicates earlier and filtering out some
long symbolic prefixes, the guided order reaches useful label-changing
branches sooner even when the final success count changes only modestly.

Overall, the sweep suggests that the method is not highly brittle to the exact \(\alpha\) value among the better-performing intermediate settings. We therefore use \(\alpha=0.6\) for the main RQ1 comparison and the RQ2 branch-ordering comparison reported above.

\paragraph{Solver search vs.~randomized input generation}
To answer the coordinate selection component of RQ3, we also checked whether pixel choice alone can explain the gain. To test this, we ran random attacks on the same two-layer Transformer, assigning random values to a pixel selected by SHAP. The success rates were in the range of 1--5\% over 100 random tests. This is far weaker than PyCT with either queue ordering or SHAP-guided ordering. This result indicates that PyCT's advantage does not come from coordinate selection alone, but depends more on symbolic solving and branch prioritization.

\rqtakeaway{RQ3}{The best performance comes from balancing decision
influence with symbolic tractability: $\alpha=0.6$ achieves the strongest
success and time tradeoff, and coordinate selection alone is insufficient
because randomized values on the same pixel selected by SHAP reach only
1--5\% success.}

\subsection{Limitations and Threats to Validity}
\label{sec:exp-limit}

Several limitations remain. First, the SHAP ranking depends on the chosen background dataset and preprocessing configuration, so the exact priority order is not fixed. Second, some reported numbers come from empirical sweeps rather than from a single frozen default. This means that the main results partially reflect parameter tuning. Third, the model family plots and the matched horizon table answer related but different questions: the plots emphasize cumulative behavior over time, whereas the table summarizes outcomes at a fixed 900\,s budget. These views are complementary, but they should not be treated as the same measurement.

Fourth, our success metric is a top-1 label change relative to the model's original prediction on the clean input, not correctness relative to ground truth. This is because both PyCT and one-pixel are applied to the same unfiltered first 100 test cases. This keeps the comparison aligned across tools, but it should be interpreted as a label change study rather than as a certified robustness evaluation on a clean and correctly classified subset. Fifth, although the CIFAR-10 inputs are normalized into \([0,1]\) before attack generation, the present experiments are reported in the model input domain and do not claim any additional realism guarantee beyond that representation.

Finally, our Transformer models are compact classifier attention networks rather than ViT encoders at full scale. We~choose this scale deliberately: the goal of the present paper is to show that concolic testing with attention semantics that are compatible with the solver is feasible, and that SHAP-guided prioritization can improve the resulting search. Scaling the same approach to larger vision transformers, richer tokenization pipelines, and stronger perturbation models remains future work.

%% file: appendix.tex
\label{app:mha-semantics}

This appendix expands the solver-compatible attention semantics in
Section~\ref{sec:transformer-semantics}. It records the tensor shapes, scalar
data flow, branch-registration points, and the concrete projection used before
\texttt{math.exp}. The implementation uses Python lists, loops, scalar
arithmetic, and ordinary conditionals so that PyCT can observe concolic branch
guards during execution.

For notation, let the input sequence be
\(X \in \mathbb{R}^{L \times d_{\mathrm{model}}}\), where row \(t\) is the
token vector at position \(t\). Let \(h\) be the number of heads and \(d_k\) the
per-head query/key dimension. The attention parameters are represented as
plain Python lists with \(W_Q,W_K,W_V \in
\mathbb{R}^{d_{\mathrm{model}}\times h\times d_k}\), \(B_Q,B_K,B_V \in
\mathbb{R}^{h\times d_k}\), \(W_O \in
\mathbb{R}^{h\times d_k\times d_{\mathrm{model}}}\), and \(B_O \in
\mathbb{R}^{d_{\mathrm{model}}}\). For concolic execution, a scalar is a pair
\(\langle c,\phi\rangle\), where \(c\) is the concrete value used in the
current execution and \(\phi\) is the symbolic expression submitted to the
solver when the operation is supported.

For head \(i\in[0,h)\), token position \(t\in[0,L)\), and feature
\(j\in[0,d_k)\), the pure-Python projections compute, for
\(T\in\{Q,K,V\}\) with matching \(W_T,B_T\),
\[
T[i,t,j] = \sum\nolimits_{k=0}^{d_{\mathrm{model}}-1}
           X[t,k] W_T[k,i,j] + B_T[i,j],
\qquad T\in\mathbb{R}^{h\times L\times d_k}.
\]
Each summation is implemented by nested Python loops, so supported arithmetic
on concolic scalars updates both the concrete value and the symbolic
expression. For each head \(i\), the score matrix
\(S_i\in\mathbb{R}^{L\times L}\) is
\[
S_i[t,u]=d_k^{-1/2}\sum\nolimits_{j=0}^{d_k-1}Q[i,t,j]K[i,u,j].
\]
This computation contains scalar arithmetic but no control-flow branch by
itself; branch predicates are introduced by the row-wise maximum used by stable
softmax.

For each row \(t\) of \(S_i\), the implementation computes
\(m_{i,t}=\max_{u\in[0,L)}S_i[t,u]\) using a Python loop of the same form as
\texttt{row\_max} in Section~\ref{sec:transformer-semantics}. Before the loop,
it calls \texttt{register\_indices} with \([(t,k)\mid k=0,\ldots,L-1]\), so the
following comparisons are associated with row \(t\) of the attention-score
matrix. When PyCT evaluates a concolic guard such as
\texttt{scores[t][u] > current}, it records the concrete branch and queues the
negation of that guard with the same path prefix; the recorded row indices let
the branch handler assign a SHAP-based priority to the corresponding
attention-row positions.

After selecting the row maximum, the row is normalized by
\[
\widehat{S}_i[t,u] = S_i[t,u] - m_{i,t},
\qquad
P_i[t,u] =
\frac{\exp\!\left(\mathrm{conc}(\widehat{S}_i[t,u])\right)}
{\sum\nolimits_{v=0}^{L-1}\exp\!\left(\mathrm{conc}(\widehat{S}_i[t,v])\right)}.
\]
Here \(\mathrm{conc}(\langle c,\phi\rangle)=c\) projects a concolic scalar to
its concrete value before the call to \texttt{math.exp}. Consequently, the
symbolic branch alternatives exposed in this softmax implementation come from
the row-maximum comparisons, while the exponential terms are evaluated with
concrete arguments. Every candidate input produced through these formulas is
still validated by re-executing the original model.

The attention output of head \(i\) and the final output projection are
\[
\begin{aligned}
A_i[t,j] &= \sum\nolimits_{u=0}^{L-1} P_i[t,u] V[i,u,j],
& A_i &\in \mathbb{R}^{L\times d_k},\\
Y[t,\ell] &= \sum\nolimits_{i=0}^{h-1}\sum\nolimits_{j=0}^{d_k-1}
  A_i[t,j] W_O[i,j,\ell] + B_O[\ell],
& Y &\in \mathbb{R}^{L\times d_{\mathrm{model}}}.
\end{aligned}
\]
This step uses nested loops and scalar arithmetic. For one head, projections
cost \(O(L\,d_{\mathrm{model}}\,d_k)\), score construction and value
aggregation each cost \(O(L^2d_k)\), and across \(h\) heads the output
projection costs \(O(L\,h\,d_k\,d_{\mathrm{model}})\). From the solver's
perspective, the relevant path predicates are the Python boolean guards
encountered during execution. The priority rule in
Algorithm~\ref{alg:concolic_execution} changes only which queued predicate is
sent to the solver first; it does not change the predicate formula or the
perturbation-domain constraints.

%% file: main.bbl
\begin{thebibliography}{10}
\providecommand{\url}[1]{\texttt{#1}}
\providecommand{\urlprefix}{URL }

\bibitem{bonaert2021fast}
Bonaert, G., Dimitrov, D.I., Baader, M., Vechev, M.: Fast and precise
  certification of transformers. In: Proc. 42nd ACM SIGPLAN Int. Conf.
  Program. Lang. Des. Implement. (PLDI), pp. 466--481. ACM, New York, NY, USA
  (2021).

\bibitem{chen2021pyct}
Chen, Y.F., Tsai, W.L., Wu, W.C., Yen, D.D., Yu, F.: {PyCT}: A Python
  concolic tester. In: Oh, H. (ed.) Program. Lang. Syst. (APLAS), Lect. Notes
  Comput. Sci., vol. 13008, pp. 38--46. Springer, Cham (2021).

\bibitem{harel2020neuron}
Harel-Canada, F., Wang, L., Gulzar, M.A., Gu, Q., Kim, M.: Is neuron coverage
  a meaningful measure for testing deep neural networks? In: Proc. ACM Joint
  Eur. Softw. Eng. Conf. Symp. Found. Softw. Eng. (ESEC/FSE), pp. 851--862.
  ACM, New York, NY, USA (2020).

\bibitem{hong2025robustdra}
Hong, C.D., Jiang, H., Lin, A.W., Markgraf, O., Parsert, J., Tan, T.:
  Extracting robust register automata from neural networks over data
  sequences. arXiv:2511.19100 (2025).

\bibitem{huang2025concolic}
Huang, M.I., Hong, C.D., Yu, F.: Concolic testing on individual fairness of
  neural network models. arXiv:2509.06864 (2025).

\bibitem{huang2020survey}
Huang, X., Kroening, D., Ruan, W., Sharp, J., Sun, Y., Thamo, E., Wu, M., Yi,
  X.: A survey of safety and trustworthiness of deep neural networks:
  verification, testing, adversarial attack and defence, and interpretability.
  Comput. Sci. Rev. \textbf{37}, Article 100270 (2020).

\bibitem{huang2023patchcensor}
Huang, Y., Ma, L., Li, Y.: {PatchCensor}: Patch robustness certification for
  transformers via exhaustive testing. ACM Trans. Softw. Eng. Methodol.
  \textbf{32}(6), Article 154 (2023).

\bibitem{jain2024towards}
Jain, S., Dutta, T.: Towards understanding and improving adversarial robustness
  of vision transformers. In: Proc. IEEE/CVF Conf. Comput. Vis. Pattern
  Recognit. (CVPR), pp. 24736--24745. IEEE (2024).

\bibitem{katz2019marabou}
Katz, G., Huang, D.A., Ibeling, D., Julian, K., Lazarus, C., Lim, R., Shah,
  P., Thakoor, S., Wu, H., Zelji{\'c}, A., Dill, D.L., Kochenderfer, M.J.,
  Barrett, C.: The {Marabou} framework for verification and analysis of deep
  neural networks. In: Dillig, I., Tasiran, S. (eds.) Comput. Aided Verif.
  (CAV), Lect. Notes Comput. Sci., vol. 11561, pp. 443--452. Springer, Cham
  (2019).

\bibitem{lin2026robustness}
Lin, L.J., Hong, C.D.: Robustness verification of recurrent neural networks
  with abstraction refinement. arXiv:2606.12490 (2026).

\bibitem{lundberg2017unified}
Lundberg, S.M., Lee, S.I.: A unified approach to interpreting model
  predictions. In: Guyon, I., von Luxburg, U., Bengio, S., Wallach, H.,
  Fergus, R., Vishwanathan, S., Garnett, R. (eds.) Adv. Neural Inf. Process.
  Syst., vol. 30, pp. 4765--4774. Curran Associates, Inc. (2017)

\bibitem{sekhon2022whitebox}
Sekhon, A., Ji, Y., Dwyer, M.B., Qi, Y.: White-box testing of NLP models with
  mask neuron coverage. In: Findings Assoc. Comput. Linguistics: NAACL,
  pp. 1547--1558. Assoc. Comput. Linguistics, Seattle, United States (2022).

\bibitem{shao2022adversarial}
Shao, R., Shi, Z., Yi, J., Chen, P.Y., Hsieh, C.J.: On the adversarial
  robustness of vision transformers. Trans. Mach. Learn. Res. (2022).
  \url{https://openreview.net/forum?id=lE7K4n1Esk}

\bibitem{shi2020robustness}
Shi, Z., Zhang, H., Chang, K.W., Huang, M., Hsieh, C.J.: Robustness
  verification for transformers. In: Int. Conf. Learn. Represent. (ICLR)
  (2020). \url{https://arxiv.org/abs/2002.06622}

\bibitem{singh2019abstract}
Singh, G., Gehr, T., P{\"u}schel, M., Vechev, M.: An abstract domain for
  certifying neural networks. Proc. ACM Program. Lang. \textbf{3}(POPL),
  Article 41, 41:1--41:30 (2019).

\bibitem{Su_2019}
Su, J., Vargas, D.V., Sakurai, K.: One pixel attack for fooling deep neural
  networks. IEEE Trans. Evol. Comput. \textbf{23}(5), 828--841 (2019).

\bibitem{sun2019deepconcolic}
Sun, Y., Huang, X., Kroening, D., Sharp, J., Hill, M., Ashmore, R.:
  {DeepConcolic}: Testing and debugging deep neural networks. In: Proc.
  IEEE/ACM Int. Conf. Softw. Eng.: Companion Proc. (ICSE-Companion),
  pp. 111--114. IEEE (2019).

\bibitem{wei2023convex}
Wei, D., Wu, H., Wu, M., Chen, P.Y., Barrett, C., Farchi, E.: Convex bounds on
  the softmax function with applications to robustness verification. In: Proc.
  Int. Conf. Artif. Intell. Statist. (AISTATS), Proc. Mach. Learn. Res.,
  vol. 206, pp. 6853--6878. PMLR (2023)

\bibitem{yu2024constraintbased}
Yu, F., Chi, Y.Y., Chen, Y.F.: Constraint-based adversarial example synthesis.
  arXiv:2406.01219 (2024).

\bibitem{zhang2024galileo}
Zhang, Y., Shen, L., Guo, S., Ji, S.: {GaLileo}: General linear relaxation
  framework for tightening robustness certification of transformers. Proc.
  AAAI Conf. Artif. Intell. \textbf{38}(19), 21797--21805 (2024).
\end{thebibliography}
